\documentstyle[aps,amsfonts,amssymb,multicol,prl]{revtex}
\input{psfig.sty}

\begin{document}

\draft
\title{
Theory of the non-equilibrium quasiparticle distribution induced by
Kondo defects}
\author{J. Kroha$^a$  and 
A. Zawadowski$^{b,c,d}$}
\address{$^a$Institut f\"ur Theorie der Kondensierten Materie, 
University of Karlsruhe, POB 6980, D-76128 Karlsruhe, Germany}
\address{$^b$Department of Physics and $^c$Research Group of the Hungarian
Academy of Sciences, Technical University of Budapest}
\address{$^d$Solid State and Optical Research Institute of the Hungarian
Academy of Sciences,  POB 49, H-1525 Budapest, Hungary}
\date{\today}
\maketitle

\begin{abstract}
It is shown that in resistive nanowires out of equilibrium containing
either single- or two-channel Kondo impurities the distribution function 
$f(E,U)$ obeys scaling behavior in terms of the quasiparticle energy $E$
and the bias voltage $U$. The numerically calculated $f(E,U)$ curves
explain quantitatively recent experiments on Cu and Au nanowires. 
The systematics of the impurity concentration $c_{imp}$ extracted from the 
comparison between theory and results on various Cu and Au samples 
strongly suggests that in these systems the scaling arises from 
magnetic Kondo impurities.
     
\end{abstract}

\begin{multicols}{2}
Electronic interactions in solids are usually investigated by means
of linear response or by spectroscopic measurements
which essentially probe the system in thermodynamic equilibrium.
In an important series of experiments \cite{pothier.97,pierre.00}
the Saclay group
demonstrated that unique information about the energy dependence
of the interactions in a mesoscopic wire can be extracted from the 
shape of the distribution function $f_x(E,U)$ of quasiparticles (qp)
with energy $E$ at a position $x$ in the wire,
when a controlled non-equilibrium situation
is established by applying a finite transport voltage $U$. 
In resistive Cu and Au nanowires out of equilibrium 
the theoretically expected double-step form of $f_x(E,U)$ \cite{nagaev.92}
was found to be rounded such that it obeys the scaling property 
$f_x(E,U)=f_x(E/eU)$, when $U$ {\it exceeds} a certain 
energy scale \cite{pothier.97,pierre.00}. By a phenomenological analysis 
within 2nd order perturbation theory, the origin of 
the scaling behavior was traced back to an
effective electron-electron interaction $v(\omega )$ which
scales with the energy 
transfer $\omega $ as $v(\omega ) \propto 1/\omega $ \cite{pothier.97}.
Such an $\omega$ dependence implies, in particular, that $v(\omega )$ 
has no essential momentum dependence and, hence, should be of local 
origin.
Moreover, within the perturbative treatment it implies a logarithmic 
divergence of the energy relaxation rate at the
Fermi energy $E_F$. The latter has generated substantial interest
because of the possible relation to the problem of
dephasing saturation \cite{mohanty.97} in mesoscopic systems. \\
\indent 
Anomalous low--energy behavior of local origin can be induced 
by the Fermi surface singularities characteristic for Kondo type systems
\cite{hewson.93,coxzawa.98}. Inelastic scattering by Kondo impurities
was discussed in \cite{zawadowski.69}.
Based on these considerations, the single--channel
Kondo (1CK) \cite{glazman.01} and the two--channel Kondo
(2CK) effect \cite{kroha.00}, possibly produced by degenerate dynamical 
defects \cite{coxzawa.98}, have been proposed as the origin of the
anomalous energy relaxation. In this Letter we 
show that a very small concentration $c_{imp}$ of either 1CK or
2CK impurities leads to the observed scaling behavior of $f_x(E,U)$,
when $eU$ exceeds an intrinsic energy scale $eU^*$ which is essentially 
equal to the Kondo temperature $T_K$. 
The numerical results are in excellent
quantitative agreement with the experimental curves 
\cite{pothier.97,pierre.00}, with $c_{imp}$ the 
only adjustable parameter of the theory. A detailed analysis suggests 
that the scaling behavior in Cu and Au wires is due to magnetic Kondo 
impurities.

Let us first set up the general formalism for calculating $f_x(E,U)$ in a 
resistive nanowire of length $L$, subject to the boundary conditions that
the left ($x=0$) and the right ($x=L$) leads are in equilibrium
at their respective chemical potentials, 
i.e.\ $f_{x=0}(E,U)=f^o(E)$, $f_{x=L}(E,U)=f^o(E+eU)$, where 
$f^o(E)=1/({\rm e}^{E/T}+1)$ is the Fermi distribution ($k_B=1$).  
The lesser ($<$) and the greater ($>$)
conduction electron Keldysh Green's functions read 
$G^<_{x} (\vec p,E) = -2\pi i f_x(\vec p)\, {\rm Im}
G^r_{x} (\vec p,E)$ and
$G^>_{x} (\vec p,E) = 2\pi i [1-f_x(\vec p)]\, {\rm Im}
G^r_{x} (\vec p,E)$, respectively, where $E$ and $\vec p$ denote energy
and momentum of the quasiparticles in a small volume centered around $x$, 
in which the external fields may be considered constant.
A superscript $^r$ indicates here and in the following a retarded
propagator. 
In a disordered electron system with diffusion constant $D$ the 
stationary quantum Boltzmann equation for the distribution as function 
of the qp energy $E$ takes the diffusive form
\cite{nagaev.92},
\begin{equation}
-D \nabla _x^2 f_x(E,U) =  {\cal C} \{ f_x(E,U) \} \ ,
\label{eq:boltzmann}
\end{equation}
The collision integral $\cal C$ is expressed in terms of the selfenergies
$\Sigma ^{\gtrless}$ for scattering into ($<$) and out of ($>$) states 
with given energy $E$ ($N_o$ = density of states per spin) as
\begin{equation}
{\cal C} = \frac{1}{2\pi N_o} \sum _p 
[ \Sigma _x ^<(E) G_{x}^>(\vec p,E) - \Sigma _x^>(E) G_{x}^<(\vec p, E) ] \ .
\label{eq:collision}
\end{equation}
In the absence of any interactions (${\cal C} \equiv 0$ 
in Eq.\ (\ref{eq:boltzmann})) the distribution
function has the double--step shape,
\begin{equation}
f_x(E,U)=\frac{x}{L} f^o(E+eU) +\Bigl(1- \frac{x}{L} \Bigr) f^o(E)\ .
\label{eq:distribution0}
\end{equation}
For a small concentration of Kondo defects $c_{imp}$, in 
addition to the static impurities, the conduction electron
selfenergy is given in terms of the single-particle   
$t$--matrix 
\begin{figure}[htb]
\centerline{\psfig{figure=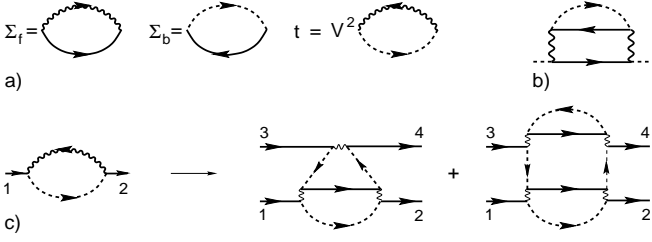,width=\hsize}}
\vspace*{0.5cm}
\narrowtext
    \caption{a) Diagrammatic representation of the NCA equations
                (\ref{eq:NCA1})--(\ref{eq:tmatrix}).
                Solid, dashed and wiggly lines denote the conduction electron,
                the local (pseudo)spin
                and the auxiliary boson propagators, respectively.
             b) Leading contribution to the inelastic spin relaxation rate 
                induced by a finite bias voltage U (see text). 
             c) Decomposition of the NCA 
                single-particle $t$--matrix $t$ to show that it includes
                the Kondo induced electron--electron vertex of $O(J^3)$
                and $O(J^4)$.}
    \label{fig:diagrams}
\end{figure}
\noindent
of the defect, $t_{x}^{\gtrless}(E)$, 
as $\Sigma_x^{\gtrless} = c_{imp} t_{x}^{\gtrless}$.
$t_{x}^{\gtrless}(E)$ mediates energy transfer between electrons in that
it couples the dynamical defect both to the in or outgoing
electron and to intermediate particle--hole pairs. 
The elastic scattering parts of $t_{x}^{\gtrless}(E)$
cancel each other exactly in $\cal C$.
We emphasize that, apart from the assumption of small $c_{imp}$,  
the present formulation, Eqs.\ (\ref{eq:boltzmann}), 
(\ref{eq:collision}), contains no approximations, 
once the $t$--matrix is known.

As pointed out in Ref.\ \cite{pothier.97}, the precise energy dependence 
of the electron-electron vertex is crucial for 
whether or not $f_x(E,U)$ obeys a 
non-equilibrium scaling property, but has been notoriously difficult to 
calculate for the Kondo problem. While 4th order (unrenormalized) 
perturbation theory yields the correct power law for
scaling \cite{glazman.01}, partial summations of 
logarithmic terms give corrections
violating scaling \cite{glazman.01,goeppert.01}.
However, such summations are valid only for $T,E \gg T_K$
\cite{hewson.93}, Chapt. 3, while the non-equilibrium situation
($eU > T_K$, $T\ll T_K$) may involve all energies $T\leq E \lesssim eU$.  

Therefore, we use the slave boson formalism, where certain exact properties 
of the auxiliary particle propagators are known 
\cite{mengemuha.88,costi.94,coxruck.93}.
To describe 1CK as well as 2CK impurities we use the
SU(N)$\times$SU(M) Anderson impurity model in the Kondo limit,
denoting the spin degeneracy by $N$ and the number of identical,
conserved conduction electron channels by $M$.  
Following the notation of Ref.\ \cite{coxruck.93}, the hamiltonian
reads,
\begin{eqnarray}
H= H_o +
\varepsilon _d \sum _{\sigma} 
f^{\dagger }_{\sigma } f^{\phantom{\dagger}}_{\sigma} 
+ V \sum _{p,m,\sigma} 
(f^{\dagger }_{\sigma} b_{\bar{m}} 
c^{\phantom{\dagger}}_{pm\sigma} + h.c.) \ ,
\label{eq:hamiltonian}
\end{eqnarray}
subject to the operator constraint
$\hat Q = \sum _{\sigma} f^{\dagger }_{\sigma } f^{\phantom{\dagger}}_{\sigma}
        + \sum _{m}  b^{\dagger }_{\bar m} b^{\phantom{\dagger}}_{\bar m} 
\equiv 1$. $H_o=\sum _{\vec p,m,\sigma}\varepsilon _p
c^{\dagger}_{\vec pm\sigma}  c^{\phantom{\dagger}}_{\vec pm\sigma}$ 
describes the conduction band. 
The auxiliary fermion and boson
operators, $f^{\dagger }_{\sigma }$, 
$b^{\dagger }_{\bar m}$, create the local defect in
its quantum state $\sigma$ or in the unoccupied state, respectively.
In the 1CK case ($N=2$, $M=1$) of a magnetic Anderson 
impurity $\sigma $ denotes spin, and $m=1$ has no relevance.
For a 2CK defect ($N=2$, $M=2$), $\sigma$ is identified with 
a pseudospin, e.g. the parity of the local defect wave
function, and $m=1,2$ is the conduction electron spin,
acting as the channel degree of freedom. 
The equilibrium Kondo temperature of the model 
is $T^{(0)}_K \simeq E_F (N N_o J)^{(M/N)} {\rm e}^{-1/(N N_o J)}$,
with $J = |V|^2/|\varepsilon_d|$ the effective spin exchange coupling.
The bare auxiliary particle propagators read
$G^{r\ (0)}_f( \omega ) =1/(\omega +i0)$ and
$G^{r\ (0)}_b( \omega ) =1/(\omega +\varepsilon _d +i0)$. Here 
we have gauged the zero of the slave particle energy
such that the pole of $G^r_f$ is at $\omega =0$.

The numerical evaluations of physical quantities will be done
within the non--crossing approximation (NCA) which is shown diagrammatically
in Fig.\ \ref{fig:diagrams} a). The corresponding equations for the
auxiliary fermion and boson
selfenergies $\Sigma _f^{\gtrless}$, $\Sigma _b^{\gtrless}$
in non-equilibrium read \cite{hettler.94}, 
\begin{eqnarray}
\Sigma _f^{\gtrless}\equiv
\frac{G^{\gtrless}_f(\omega)}{|G^{r}_{f}(\omega)|^{2}} &=& 
   - \frac{M \Gamma}{N_o}  \int \frac{d\varepsilon}{2\pi i}\, 
     G^{\gtrless}_{x}\, (- \varepsilon ) 
     G^{\gtrless}_b(\omega + \varepsilon) 
\label{eq:NCA1} \\
\Sigma _b^{\gtrless}\equiv
\frac{G^{\gtrless}_b(\omega)}{|G^{r}_{b}(\omega)|^{2}} &=& 
   + \frac{N \Gamma} {N_o} \int \frac{d\varepsilon}{2\pi i}\, 
     G^{\lessgtr}_{x}\, (  \varepsilon ) 
     G^{\gtrless}_f(\omega + \varepsilon) \ ,
\label{eq:NCA2}
\end{eqnarray}
where $\Gamma = \pi N_o|V|^2$ is the effective hybridization,
and $G^{\gtrless}_{x}\, (  \varepsilon ) = 
\sum _p G^{\gtrless}_{x}\, (\vec p,\varepsilon )$.  
This set of selfconsistent, non--linear equations is closed by the 
Kramers-Kroenig relations, 
$G_{f,b}^{r}(\omega ) = - \int d\varepsilon/(2\pi i)\, G_{f,b}^{<}(\omega )/
(\omega-\varepsilon+i0)$, which follow from causality and the fact
that the auxiliary particle Green's functions
have only forward in time propagating parts.
Within NCA
the single--electron $t$--matrix due to the Kondo impurity is
\begin{eqnarray}
t_{x}^{\gtrless}(E ) =  
   - \frac{\Gamma}{\pi N_o} \int \frac{d\varepsilon}{2\pi i}\, 
     G^{\gtrless}_{f}\, (E +  \varepsilon ) 
     G^{\lessgtr}_b(\varepsilon) \ . 
\label{eq:tmatrix}
\end{eqnarray}
By writing the renormalized auxiliary propagators as the bare ones
with selfenergy insertions and using $|V|^2G_b^{r (0)}(\omega \approx 0)=J$, 
it is seen that the present formulation includes the Kondo induced 
electron-electron vertex  
of $O(J^3)$ and $O(J^4)$, with the points 3 and 4 connected 
(Fig.\ \ref{fig:diagrams} c)). The exchange diagrams (points 3 and 2 
connected) are not included. However, this does not change the
scaling properties (see below). By self-consistency, NCA goes beyond this
two-particle scattering approximation considered in 
Ref.\ \cite{glazman.01}. 

The NCA in non-equilibrium includes both an inelastic spin 
relaxation rate and scaling behavior in terms of the applied bias.
It is instructive to investigate these properties  
for a single Kondo impurity before we present the numerical solutions.
The Kondo scale $T_K$ is influenced by the step heights in $f_x(E,U)$,
Eq.\ (\ref{eq:distribution0}),
and the size of the corresponding logarithmic terms in the
Kondo vertex, which is reduced compared to equilibrium. 
This leads to a suppression of $T_K$, e.g. in the middle of the 
wire ($x/L = 1/2$), 
\begin{equation}   
T_K = \sqrt{(eU/2)^2 + T_{K}^{(0)\ 2}}-eU/2 \stackrel{eU\gg T_{K}^{(0)}} 
{\simeq} 
\frac{T_{K}^{(0)\ 2}}{eU} \ . 
\label{eq:TKeU}
\end{equation}   
At an arbitrary position $x/L$, for $eU \gg T_K^{(0)}$ we have
$T_K = T_K^{(0)\ 1/\eta }/(eU)^{(1/\eta) -1}$, where
$\eta = {\rm max} [x/L, 1-x/L]$.
At the same time, there is
an inelastic spin relaxation rate $1/\tau _s$,
since in the non-equilibrium electron sea
(Eq.\ (\ref{eq:distribution0})) there is finite
phase space available for scattering even at $T=0$. Technically, this
relaxation rate appears as the imaginary part of the pseudofermion
selfenergy, $\Sigma _f^r(\omega=0)$, 
which carries the local spin degree of freedom.
To leading order in $J$ it is obtained by inserting the bare propagators
$G_{f,b}^{\gtrless ^{(0)}}$ in the diagram Fig.\ \ref{fig:diagrams} b), 
\begin{eqnarray}
\frac{1}{\tau_s} =  
2\pi M N \frac{x}{L} \Bigl( 1 - \frac{x}{L}\Bigr) (N_o J)^2 eU \ .
\label{eq:taueU0}
\end{eqnarray}
This is analogous to the well--known Korringa spin relaxation rate
\cite{hewson.93}, with $T$ replaced by $eU$. 
Solving Eqs.\ (\ref{eq:NCA1}), (\ref{eq:NCA2}) selfconsistently in the
complete range of validity of NCA, $T_K \lesssim eU \ll E_F$, we find
that beyond perturbation theory $1/\tau_s$ depends on $eU$ and $T_K$ only,
\begin{eqnarray}
\frac{1}{\tau_s} =  
 \frac{x}{L} \Bigl( 1 - \frac{x}{L}\Bigr) H_{M,N}\Bigl(
\frac{eU}{T_K}\Bigr) eU \ ,
\label{eq:taueU}
\end{eqnarray}
where the universal function $H_{M,N}(y) \to \pi M/[2N{\rm ln}^2 (y)]$
for $y\gg 1$
\cite{kroha.01}, in accordance with Ref.\ \cite{glazman.01}. 
Inserting $1/\tau _s$ into the pseudoparticle propagators, 
it cuts off all logarithmic contributions of perturbation theory. Thus, the
low--$T$ scale of the non-equilibrium Kondo system is
$T_o = {\rm max} [T_K(eU), 1/2\tau _s (eU)]$.
The crossover from the Kondo (Eq.\ (\ref{eq:TKeU})) or $T$
limited life time to the inelastic scale
(Eq.\ (\ref{eq:taueU})) occurs as function of $eU$ at 
a bias $eU^*$. It follows from the universality of $1/\tau _s$ that 
$eU^*$ is only a function of $T_K^{(0)}$, i.e., for dimensional reasons,
$eU^* = A_{M,N} T_K^{(0)}$. Numerically we find $A_{1,2}= 1.48\pm 0.08$ 
and $A_{2,2}= 1.39\pm 0.05$. 
For $eU >eU^*$ $T_K$ has lost its relevance, and for
$eU \gtrsim 10 eU^*$ one has with good accuracy $1/\tau _s \propto eU$,
when $eU$ is varied by a factor of $\sim 4$, wherein the $eU$ dependence
of the log terms is weak.   

To investigate scaling of $f_x(E,U)$ we must 
consider the energy dependence of the exact pseudoparticle Green's functions,
$G_{f,b}^{\gtrless}(\omega)$, 
from which all other physical quantities are derived.
It is known that in equilibrium at $T=0$ it is
determined by an infinite logarithmic series which results in
power law behavior, $G_f^{\gtrless}(\omega) 
\propto i \Theta ( \pm \omega )|\omega |^{-\alpha _f}$,
$G_b^{\gtrless}(\omega) \propto i\Theta ( \pm \omega )|\omega |  
^{-\alpha _b}$ for $\omega \lesssim T_K^{(0)}$. 
The exponents $\alpha_f$, $\alpha_b$ are due to
an orthogonality catastropy in the auxiliary propagators and have 
characteristic values 
$\alpha _f = \alpha _b = 1/2$ for the 1CK and 
$\alpha _f = M/(M+N)$, $\alpha _b = N/(M+N)$ for the
2CK fixed point of the model Eq.\ (\ref{eq:hamiltonian}) (Kondo limit) 
\cite{costi.94,coxruck.93}. We can exploit this knowledge to 
determine the frequency dependence of $G_{f,b}^{\gtrless}(\omega)$ 
away from equilibrium without explicitly summing up the logarithmic series.
At finite bias $eU \gg T_K$ 
this series consists of similar terms as in equilibrium,
however with three modifications:
($i$) Because of the inelastic relaxation rate $1/\tau _s =2\gamma $ 
all frequency arguments are shifted, $\omega = \omega + i \gamma$.
($ii$) $G_f(\omega )$ has a singularity at 
$\omega = 0 + i\gamma$, 
but there are two singularities in $G_b^{\gtrless}(\omega )$ at
$\omega = 0 + i b \gamma $ and at $\omega = eU + i b \gamma $, where
$b$ is a numerical factor.
($iii$) Each frequency integral 
involving $G_b^{\gtrless} (\varepsilon + \omega )$, like, e.g., 
in Eq.\ (\ref{eq:NCA1}), carries a prefactor $M$, and
each of the two singularities in $G_b^{\gtrless}$ gives a singular
contribution of equal weight at the external frequency $\omega=0$. 
This can be seen as an effective doubling of $M$.  
Points ($i$)--($iii$) can be verified by iterating 
Eqs.\ (\ref{eq:NCA1}), (\ref{eq:NCA2}), starting from the bare propagators
$G_{f,b}^{\gtrless\ (0)}$. 
As a result, we obtain at $x/L = 1/2$ 
damped power law behavior for the
auxiliary propagators in non--equilibrium,
\begin{eqnarray}
G_f^{>}(\omega ) &\propto&   i {\rm Im} 
\frac {1-f_{1/2}(E,U)}{(\omega + i \gamma )^{\alpha _f'}} 
\label{eq:Gf_noneq} \\ 
G_b^{>}(\omega ) &\propto&  i {\rm Im}\Bigl[ 
\frac {1-f_{1/2}(E,U)}{(\omega + i b \gamma )^{\alpha _b'}} +
\frac {1-f_{1/2}(E,U)}{(\omega -eU + i b \gamma)^{\alpha _b'}} 
 \Bigr] \ .
\nonumber
\end{eqnarray}
The exact exponents \cite{coxruck.93}
in the non-equilibrium situation (with $M \to 2M$ in the 
logarithmic series) are
$\alpha _f' = 2M/(2M+N)$, $\alpha_b' = N/(2M+N)$.
The $\omega$ dependence Eq.\ (\ref{eq:Gf_noneq}) 
should extend from $\omega =0$ up to the smallest energy scale of the model,
i.e. for $eU > eU^*$ up to $\omega = eU$, since in this case  
the Kondo scale has disappeared. The behavior described above is
confirmed by our numerical NCA solutions. For $x/L \to 0$ or  $x/L \to 1$
the solution crosses over to the equilibrium one, as expected.
The modification of 
the exponents $\alpha _f'$, $\alpha _b'$ compared to their
equilibrium values is reminiscent of a doubling of the channel number due to
the two Fermi edges. It remains to be seen whether a 
strong coupling region ($T_K^{(0)}<eU<eU^*$) can be realized where 
such behavior can be observed
in the presence of $1/\tau _s \simeq O(eU)$. The latter was neglected in 
Ref.\ \cite{coleman.00}. Here we are interested in scaling at large bias
($eU \gg eU^*$). Inserting the power law 
forms Eq.\ (\ref{eq:Gf_noneq}) 
into Eqs.\ (\ref{eq:NCA1})--(\ref{eq:tmatrix}),
dividing Eq.\ (\ref{eq:NCA1}) by $(eU)^{\alpha _f'}$ and 
Eq.\ (\ref{eq:NCA2}) by $(eU)^{\alpha _b'}$, and using the exact result
$\alpha _f' + \alpha _b' =1$, it is seen that the NCA equations contain 
only dimensionless energies, $\varepsilon /eU$ etc. Power counting 
arguments \cite{coxruck.93} show that 
this is reproduced in arbitrary selfconsistent order in $\Gamma$ beyond NCA. 
In the presence of a finite concentration
$c_{imp}$, $f_x(E,U)$ is determined by the selfconsistent 
coupled set of equations (\ref{eq:boltzmann}), (\ref{eq:collision}) 
and (\ref{eq:NCA1})--(\ref{eq:tmatrix}). 
It follows that the solution obeys scaling, $f_x(E,U)=f_x(E/eU)$  for
$eU > eU^*$. Our numerical solutions show scaling within a 
factor of 4 to 9 in $eU$, depending on parameters, wherein log corrections
to $1/\tau_s \propto eU$, Eq.\ (\ref{eq:taueU}), are small. 
Note that the power law behavior Eq.\ (\ref{eq:Gf_noneq})
and the fact that the low-energy cutoff $1/\tau _s$ itself is proportional
to $eU$ (up to small log corrections) 
cooperate to produce scaling. For $eU \lesssim 10 eU^*$ we find
deviations from scaling, because then the latter condition is no longer
fulfilled. This provides for $T \ll T_K^{(0)}$ a rough
estimate, and for $T > T_K^{(0)}$ an upper bound on $T_K^{(0)}$;
in the experiments \cite{pothier.97,pierre.00} $T \lesssim T_K^{(0)} \ll eU$. 

For the numerical evaluations we assume magnetic 
\end{multicols}
\widetext
\begin{figure*}
\centerline{\psfig{figure=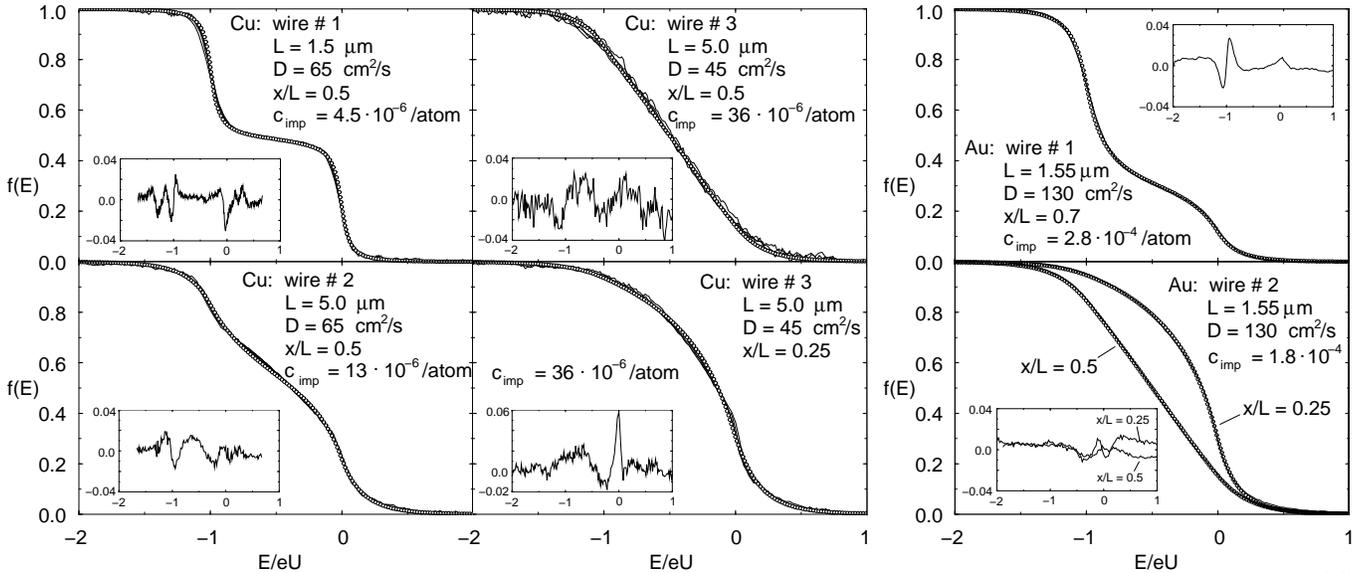,width=\linewidth}}
\vspace*{0.00cm}
    \caption{Non--equilibrium distribution functions for various Cu and
      Au samples. Black lines: experimental results; Cu: [1],
      Au: [2]. Open circles: theory for $eU \gg eU^*$. 
      Deviations from scaling at smaller $eU$ [1]
      are also reproduced by the theory (not shown).
      Fitted $c_{imp}$ values are indicated. The insets
      show the difference between the experimental and the theoretical curves.
       }
    \label{fig:distribution}
\end{figure*}
\begin{multicols}{2}
\narrowtext
\noindent
 (1CK) impurities
(2CK impurities give very similar results) and take 
$T_K^{(0)} \approx 0.1$ K in Cu  and 
$T_K^{(0)} \approx 0.5$ K 
in Au wires (corresponding to $N_oJ = 0.041$
and $N_oJ = 0.048$, respectively),
consistent with the above estimate and with 
independent estimates of $T_K^{(0)}$ for these samples \cite{pierre.00}. 
After $T_K^{(0)}$ is fixed, $c_{imp}$ is the
only adjustable parameter of the theory.
The results for $f_x(E,U)$, as measured by a tunnel junction
attached to the wire, are shown in Fig.\ \ref{fig:distribution}.
Excellent quantitative agreement with experiments  
\cite{pothier.97,pierre.00} is obtained for all samples.
In Au wires the fitted values of $c_{imp}$ are 
consistent with  (although somewhat higher than)
independent estimates of the magnetic impurity
concentration \cite{pierre.00}, considering the roughness of both
estimates. This suggets that the scaling behavior of $f_x(E,U)$ in
the Au samples is due to magnetic (1CK) impurities. Furthermore,
in all Cu samples the fitted $c_{imp}$ is $\sim 10^{2}$ times smaller 
than in Au. This systematics is in accordance with $c_{imp}$ estimated
from the plateau in the $T$ dependence of the dephasing time 
$\tau _{\varphi}$ in similarly prepared samples \cite{pierre.00,gougam.00}.\\
\indent
In conclusion we have shown that single-- or two--channel Kondo impurities
in quantum nanowires 
induce scaling behavior of the non-equilibrium distribution function 
$f_x(E,U)$ at a bias $eU$ exceeding an
energy scale $eU^*\approx T_K^{(0)}$. 
The results give a detailed 
explanation of related experiments. In the small bias or strong coupling 
regime ($T_K>eU$), 1CK and 2CK impurities must show qualitatively 
different behavior, as the former become potential scatterers with
frozen spin dynamics, contrary to the latter with (ideally) non--zero
entropy at $T=0$. The quantitative comparison between the present 
theory and experiments suggests that in 
Au and at least partially in Cu nanowires both the scaling
of $f_x(E,U)$ \cite{pothier.97,pierre.00} and the plateau in  the
low--$T$ dephasing time $\tau _{\varphi}$ \cite{gougam.00}  
are due to magnetic Kondo impurities. A unique test for 
magnetic impurities will be measuring $f_x(E,U)$ in a magnetic field. 

We are grateful to H.~Pothier, B.~Al'tshuler, N.~O.~Birge, 
J.~v.~Delft,
M.~Devoret, D.~Esteve, A.~Rosch, and P.~W\"olfle for helpful 
discussions.
This work was supported by DFG through SFB195, by Hungarian  grants
OTKA T024005, T029813, T034243 and by the A.~v.~Humboldt foundation.

\end{multicols}

\end{document}